# A Framework for Detection and Classification of Attacks on Surveillance Cameras under IoT Networks

Umair Amjid, M. Umar Khan, S. A. Manan Kirmani
Department of Computer Engineering,
COMSATS University Islamabad, Islamabad, Pakistan.
umar_khan@comsats.edu.pk

## ABSTRACT

The increasing use of Internet of Things (IoT) devices has led to a rise in security related concerns regarding IoT Networks. The surveillance cameras in IoT networks are vulnerable to security threats such as brute force and zero-day attacks which can lead to unauthorized access by hackers and potential spying on the user's activities. Moreover, these cameras can be targeted by Denial of Service (DOS) attacks, which will make it unavailable for the user. The proposed AI based framework will leverage machine learning algorithms to analyze network traffic and detect anomalous behavior, allowing for quick detection and response to potential intrusions. The framework will be trained and evaluated using real-world datasets to learn from past security incidents and improve its ability to detect potential intrusion.

## 1.1 Introduction

The existing digital context has no doubt identified the emergence of Internet of Things (IoT) devices as the main reason for the improvements of the comfort and efficiency of human lives. Now days from smart thermostats to wearable fitness tracker these interconnected devices have been smoothly incorporated in as similar to our daily activities. However, with this exponential growth in IoT adoption comes a pressing concern: protection of privacy, data breaches, and security weaknesses are the key within e-commerce.

Thereby, IoT devices are created for the purpose of getting, sending and even analyzing data on their own. This gives rise to a situation where both, convenience and automation are expanded, but also rich in a Pandora's box of opportunities for perils. Because hackers are now searching for the security holes to steal the data or just to cause harm, they are targeting the IOT devices as their main objects of hacking.

It may be convenient that the insecure IoT devices allow ease of access, but at the same time, it introduces the risk that unauthorized individuals could gain access. Take a breather. Now assume that an intruder manages to take over the smart surveillance camera in your premises via a brute force or zero-day attack. Now, what was until then confidential is opened to the public and if hackers decided to take a chance on your moves, they would have no trouble.

In addition to that, IoT devices are not shielded against the type of Distributed Denial of Service (DDoS) attacks, in which hackers attack the server or a network with so much traffic that it becomes unavailable for all the legitimate users. This brings us to the question whether the deteriorating market security has a devastating impact, especially in emergency when the order of working of IoT devices matters.

There are security threats that unfavorably affect individuals' privacy beyond the privacy issues. From things like healthcare, where IoT devices are used more and more for remote patient monitoring and medical diagnostics and breach in security of which may lead to death, the consequences of successful intrusion are critical.

To deal with this security maze there must be a multi-faceted approach. The security must be originally considered during the design and production of IoT devices, the development should continue with incorporating potent encryption protocols, regular security updates and strict access controls. Additionally, consumers should be cautious as well, make it a habit to update their IoT devices regularly and ensure the use of unique, strong password for them.

Moreover, government authority and regulators are at the top to set IoT industry standards and are aimed at protection against IoT security threats. Through the building of an association between all the key players in the sector and the promotion of safe cybersecurity practices, we will be able to deal with inactive devices and at the same time harness and realize the full potential of the Internet of Things in a secure and connected society.

## 1.2 Motivation

The core reason is the implementation of stronger cybersecurity mechanisms to avoid access to users' privacy and security. The modern-day internet of things (IoT) cameras is an effective tool for surveillance and monitoring. If this is compromised, it can lead to unintended breaches of privacy that can also worsen security at physical locations.

Through detection and classification of a IoT cameras assaults connected to cyber threats, the Project is designed to strengthen security against cyber-attacks on camera IoT.

In addition, this project gives a remedy to the surplus problems associated with security flaws of IoT devices. In addition to the privacy concerns of individual people, cameras that are compromised would then be a tool in the network of large cyberattacks, thus co-operating with botnets and espionage. The goal of the project is to prevent attacks on IoT cameras by identifying threats and finding effective ways of dealing with them. The project thus aims to safeguard the individual users as well as the overall integrity of IoT networks and other related wired things.

In addition, the project brings an anticipatory approach into the picture which reacts to the fact that the danger environment is changing all the time. With cyber threats organizing and developing in an advanced and ubiquitous way, it is important that the developed solutions related to security are constantly evolved toward maintaining their security high and returnable. Through carrying out the work which aims to discover and sort hacker efforts on IoT cameras this project would develop cybersecurity technologies using various technologies thus making any IoT space free of any attacks.

Basically, the motive of the work is to resolve the place of research on the surveillance cameras security in the area of internet of things devices, which are the vital part of the modern cities. The project intends to achieve this by monitoring and characterizing attacks on such equipment allowing for privacy safeguard against cyber threats and contributing to the fight against rising in Internet connected term.

## 1.3 Problem Statement

In the modern world, high density of security cameras has accentuated because of their widespread nature. They are not only deployed for observing and monitoring indoor spaces like homes, but also for many other purposes like public spaces, critical infrastructure and so on. Nevertheless, these wide spreads of deployment have made security cameras easily accessible to them who are seeking for the weak spots to enjoy the shortcoming for their negative objectives. There's been a growing number of incidents where security cameras have been crippled by a cyber-attack. This not only shows the potential risks to individuals' privacy but to also incite a danger to property and public safety.

Despite all the attempts that have been made to strengthen the security systems, it still seems that the present arrangements are generally not in a position to give adequate reaction to this kind of an attack. Through traditional security platforms including firewalls and antiviruses, you may reach some level of security guards; however, these tactics are still not usually capable to withstand successful cyber-attacks. This hence demands new and more effective solutions to respond to current events and detect and handle impending risks.

In order to apprehend this challenge, the project will be designed with an AI model that is a special type of AI model developed to make security surveillance system protected. Our approach involves expenditure of AI technologies, including machine learning and deep learning, to analyses network traffic and other types of relevant data which are abnormal and similar to the typic current of attacks. Unlike rule-based systems as they operate on predetermined criteria and are thus limited as regards their variability of adaptation an AI program will be gradually be acquiring new info while remaining to advance by itself. Thus, it will stay ahead of emerging looms.

Our approach root is due to the fact that it is capable to detect many types of attacks precisely including but not restricted to denial-of-service (DoS) attacks, malware injections, and unauthorized access attempts. The AI model we create will allow during transmissions and this will give the guards an edge to act fast and curb any emerging issues on stream. Therefore, this will not only enhance the reliability of the surveillance systems but also guarantee their integrity.

## 1.4 Outline of Approach

First the data is gathered form CIC (Canadian institute of cyber security). The attack dataset generated by them in case of brute force and Dos attacks are considered. Then python is used with different libraries such as pyshark**[24]**, Tshark**[22]**, Nfstream**[23].** Using these libraries the useful feature is extracted such as ttl, port number, source port, destination port, Ip address and so on and so forth.

| Model Evaluation Metrics | Pyshark | Tshark | Nfstream |
|---|---|---|---|
| Number of features | 19 | 28 | 27 |
| Complexity in coding | High | Low | Low |
| Speed and efficiency during execution | Very Slow | Very Fast | Fast |
| Results in training the model | Excellent | Excellent | Poor |

## 1.4.1 Machine Learning

These activities become the means for extracting and inspecting the data with NumPy and pandas to have a clear-cut picture of the dataset's architecture, distributions, and how the variables relate to each other. Exploratory Data Analysis (EDA) approaches using summary statistics, plots, and correlational analysis are also used to draw out patterns and insights relevant to feature selection and model building.

The data is preprocessed after EDA to deal with missing values, encode variables that are categorical and scale the features as a rule in order to make it available for the machine learning models. This preprocessing step ensures that the data is in the proper framework for modelling and also it helps to mitigate issues like data imbalance and feature skewness problems.

The model training cycle comprises of the assessment of the various machine learning algorithms namely the Random Forest, AdaBoost and K-Nearest Neighbors (KNN) along with the Gaussian Naive Bayes. Each method is assessed with the help of techniques such as cross validation and hyperparameter tuning in order to achieve maximum metrics such as correctness, exactness, sensitivity, and F1 score. This is an important step to deduce what type of algorithm that fits well with the particular dataset and a particular problem that the algorithm is supposed to solve.

To ensure the models' calibration and the predictive effects, the trained models are tested using different dataset sizes, paying a special attention to bigger datasets. Through this process of testing the presence of anomalies such as overfitting is avoided, where the performance of the model is good with the training data, yet the performance is not good on the unseen data. To resolve various kinds of overfitting, methods like regularization, feature selection, or increasing the number of cases in training data might be helpful.

## 1.4.1.1 Different models used for Training
i.    Random Forest
ii.    Navie Bayes
iii.    Ada-boost
iv.    Perceptron

Random Forest:

Random Forest is a synchronous ensemble approach applied for both classification and regression problems. It is based on tree-based models, and the outputs of several trees are aggregated to increase accuracy and generalization. All decision trees in the forest are being trained on a random chunk of the data and equipped with a different set of features to fight overfitting and keep diversity. The last phase is called prediction during which the algorithm adds up the prediction of all the trees and determines the final output either via averaging or a majority mechanism. The advantage of Random Forest is its resistibility to the noise marker data, its usefulness in a high-dimensional space, and its ability to show conclusions on the features - therefore its frequent use in different machine learning applications is not surprising.

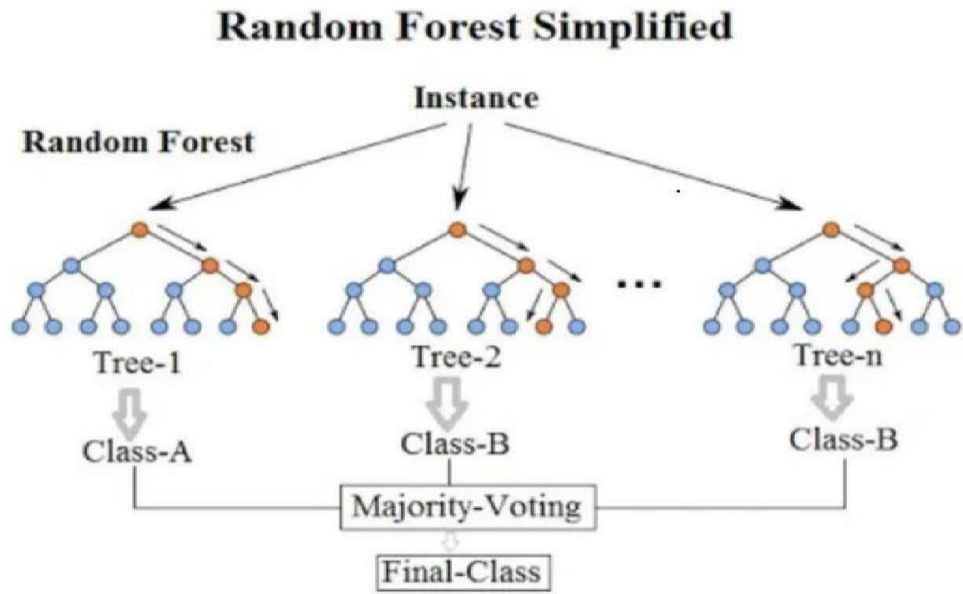

*Figure 1. 1: Will Koehrsen, 2017 Random Forest Simplified.*
*[online image]*

Navie Bayes:

Naive Bayes is a probabilistic classification algorithm based on Bayes' theorem, with the "naive" assumption of independence among features. It calculates the probability of a given instance belonging to a particular class by multiplying the probabilities of individual features occurring in that class and normalizing by the overall probability of those features occurring. Despite its simplicity and assumption of feature independence, Naive Bayes often performs surprisingly well in practice, especially for text classification tasks.

Ada-boost:

Ada-boost, actually the short form of Adaptive Boosting, is a machine learning algorithm which adds the outputs of weak learners to the output of a complicated learner. The poor learners are very straightforward and simple equipped with decision trees limited for the generalization. Ada-boost does so sequentially put emphasis on the weak learners before training starts with subsets of the data while giving higher weights to instances that were misclassified by earlier models. At every stage, the algorithm corrects the weights of incorrect classifications, thus one of its next steps is precisely this. The model finally is a measured combination of outputs of the weak learners which are the base predictions where each learner's weight is determined by its accuracy level. AdaBoost construction is very proficient in the tasks of binary classification, that it can get high precision and deal with overfitting well.

Perception:

The perceptron is the basis and the fundamental algorithm in the area of neuronal networks and the supervised learning, which mostly is the base of more complicated models. It is a two-envelope feature learner, which learns the weights of those features for decision-making. A perceptron receives many input values, of which each one is a product of its multiplying by its particular weight and the sum of all these products. This leads to calculated values next being received by an activation function that uses a step function for binary signals to produce an output. The weights of the perceptron which is a function of the errors between the actual labels and the predictions converge to a conditional independence property (the margin).

## 1.5 Main Contribution

The work has significantly advanced by performing real testing of an IoT cameras' breaches and with a capability unfortunately not so often demonstrated in the round of studies in the field. This achievement indicated an important stride where the project highlighted its innovation again and its focus on dealing with the IoT devices' security concerns.

By organizing this introducing live tests with special attention paid to various sorts of attacks purposely aiming on IoT cameras, the project crew managed to do it. This research was performed outdoors, exactly as it would be in the case of an attack with the evaluation of results from it valid and trustworthy. This attack simulation on the IoT cameras, not only exposed their weaknesses but also provided objective ground to back their success in survival of these threats.

The impressive level of accuracy during the implement phase of the project highlights the high performance of the methods and technologies used by the project. Such certifications reflect the project team's professionalism and proactiveness, and also confirm their resolve to protect the interconnectedness and credibility of IoT.

However, IoT security research and development cannot be overlooked after the live test has revealed successful results. The outcomes of these tests give significant data and observations that can be used for continuous improvements on the current security metrics or can be the source of novel ideas that can tackle new threats being discovered.

Through the live testing of attacks on the IoT cameras that are conducted with a significant level of precision, the project as improved academic know-how and can be influential in guiding industry practices and frame policy decisions. The result of the analysis will be implemented throughout the Internet of Things systems, hence, leading to a global level of security improvement. Furthermore, the digital world will be safer and more secure.

# 1.6 United Nation Sustainable Development Goals (UN SDGs)

This work is in line with the following UN SGDs. Detections and Classification of Attacks on Surveillance Cameras contribute to several Sustainable Development Goals (SDGs) through their application in various fields.

- **SDG 9: Industry, Innovation, and Infrastructure:** This project aligns with SDG 9 by contributing to the development and enhancement of surveillance technology and infrastructure. By detecting and classifying attacks on surveillance cameras, it supports the goal of building resilient infrastructure, promoting inclusive and sustainable industrialization, and fostering innovation.

- **SDG 16: Peace, Justice, and Strong Institutions:** Surveillance cameras play a crucial role in maintaining peace, ensuring justice, and strengthening institutions by providing evidence and deterring crime. Detecting and classifying attacks on surveillance cameras can help in ensuring the effectiveness and reliability of these systems, thereby contributing to the promotion of peaceful and inclusive societies for sustainable development.

- **SDG 11: Sustainable Cities and Communities:** The project supports SDG 11 by enhancing the security and safety of urban areas through the protection of surveillance infrastructure. By detecting and classifying attacks on surveillance cameras, it contributes to creating more resilient and sustainable cities, ensuring the well-being and safety of residents.

# 2.1 Literature Review

This paper examines the fusion of Internet of Things (IoT) security and machine learning techniques. It investigates the utilization of artificial intelligence (AI) by IoT devices to fortify their security protocols. By integrating AI algorithms, IoT devices can bolster their capabilities in identifying and mitigating security threats effectively. This symbiosis of AI and IoT technology enhances the overall security infrastructure within IoT ecosystems, ensuring a more robust defense mechanism against potential cyber risks. Ultimately, this innovative approach highlights the potential of leveraging machine learning for IoT security to create a safer interconnected environment.**[1]**

Smart cities utilize IoT technology for urban informatization, but security risks arise from wireless sensor network devices. LoRa communication technology can improve security in smart cities. A new backoff algorithm, REBEB, improves security performance and fairness, with fairness reaching 0.985 and better when nodes are small.**[2]**

This survey paper reviews the application of artificial intelligence (AI) techniques to enhance the security of Internet of Things (IoT) systems. The authors explore various AI methods, including machine learning, deep learning, and reinforcement learning, and their roles in addressing IoT security challenges such as intrusion detection, anomaly detection, and threat mitigation. The paper highlights the advantages of using AI to provide real-time, adaptive security solutions and discusses the current state of AI-driven IoT security research. The authors also identify key challenges and propose future research directions to further improve AI-based IoT security frameworks.**[3]**

The paper "Research on AI Security Enhanced Encryption Algorithm of Autonomous IoT Systems" presents an AI-driven encryption algorithm to improve security in IoT systems. Traditional encryption methods are inadequate for dynamic IoT environments, so the proposed solution integrates AI to dynamically adapt and respond to threats. The algorithm shows superior performance in detecting and mitigating security risks in real-time, enhancing the overall security without significant overhead. The results indicate improved resistance to attacks and efficient encryption processes. This research highlights the potential of AI-enhanced encryption to protect IoT data and suggests future optimization and application exploration.**[4]**

This paper presents methods to improve authentication and authorization processes in large-scale IoT systems using AI techniques. The proposed methods aim to speed up authentication and provide a scalable solution for progressive authorization. AI algorithms are leveraged to identify and respond to potential security threats dynamically. The study demonstrates how AI can significantly enhance security measures, ensuring rapid and secure access control in extensive IoT networks.**[5]**

The authors propose a comprehensive security framework for IoT systems using machine learning techniques. The framework addresses various security challenges such as intrusion detection, anomaly detection, and threat prediction. By integrating machine learning models, the framework can dynamically adapt to new threats and improve the overall security posture of IoT systems. The paper highlights the effectiveness of machine learning in enhancing IoT security through real-time threat analysis and response.**[6]**

This review article analyzes the security challenges faced by IoT systems and explores AI-based solutions to address these issues. The authors discuss various security threats, including data breaches, unauthorized access, and malware attacks. They propose using AI techniques like machine learning, deep learning, and reinforcement learning to enhance security measures. The paper concludes that AI can significantly improve IoT security by providing adaptive and intelligent threat detection and mitigation strategies.**[7]**

This comprehensive overview examines the current state of IoT security and explores machine learning solutions to enhance security intelligence. The authors review various machine learning algorithms and their applications in detecting and mitigating IoT security threats. They identify key research directions for improving IoT security using AI, including the development of more robust and scalable machine learning models. The paper emphasizes the importance of integrating machine learning into IoT security frameworks for better threat detection and prevention.**[8]**

This paper introduces a distributed architecture for evaluating AI-based security systems at the network edge using the TON_IoT datasets. The proposed architecture aims to provide a scalable and efficient solution for assessing the performance of AIdriven security measures. The authors demonstrate how the architecture can be used to test various AI algorithms for intrusion detection, anomaly detection, and other security applications. The study highlights the importance of edge computing in enhancing the security of IoT systems.**[9]**

The authors propose a two-stage intrusion detection system for software-defined IoT networks, leveraging AI techniques. The first stage uses machine learning algorithms for initial anomaly detection, while the second stage employs deep learning models for precise intrusion classification. The system is designed to improve detection accuracy and reduce false positives. Experimental results show that the proposed system outperforms traditional intrusion detection methods in terms of both accuracy and efficiency.**[10]**

This review article provides an overview of AI-driven cybersecurity, focusing on security intelligence modeling and future research directions. The authors discuss various AI techniques, including machine learning, deep learning, and reinforcement learning, and their applications in cybersecurity. The paper highlights the benefits of using AI for real-time threat detection, predictive analytics, and automated response. It also outlines key challenges and opportunities for advancing AI-driven cybersecurity research.**[11]**

This paper explores the use of deep learning algorithms to classify IoT security risks. The authors develop a deep learning-based model to identify and categorize different types of security threats in IoT systems. The model is trained on a large dataset of IoT security incidents and demonstrates high accuracy in threat classification. The study shows that deep learning can effectively enhance IoT security by providing detailed and accurate threat analysis.**[12]**

The authors propose AI-based architectures for sustainable IoT security. They discuss the integration of AI techniques like machine learning and deep learning into IoT security frameworks to ensure long-term sustainability. The paper highlights the benefits of AI in enhancing security, such as improved threat detection and adaptive responses. The authors also address the challenges of implementing AI-based security solutions in IoT environments and suggest potential research directions.**[13]**

This research article presents a machine learning-based approach to detect and mitigate denial of service (DoS) attacks in IoT systems. The authors introduce a "looking-back" technique that leverages historical data to enhance the accuracy of machine learning models in identifying DoS attacks. The proposed method shows significant improvements in detection rates and mitigation effectiveness compared to traditional techniques. The study emphasizes the importance of using machine learning for proactive IoT security.**[14]**

This comprehensive survey examines the application of reinforcement learning (RL) in IoT security. The authors review various RL algorithms and their use cases in enhancing IoT security measures. The paper discusses the benefits of RL, such as adaptive learning and decision-making capabilities, in responding to dynamic security threats. The authors also identify challenges and future research directions for integrating RL into IoT security frameworks.**[15]**

Security cameras are essential for safeguarding various environments, including residential, commercial, and public spaces. However, the increasing interconnectedness of devices and the Internet of Things (IoT) expose security cameras to cyber-attacks, posing risks to privacy, property, and public safety**.[16]**

This project aims to address these emerging threats by leveraging artificial intelligence (AI) models to enhance the security and resilience of surveillance systems. By analyzing network traffic patterns, the project seeks to accurately detect and mitigate attacks on security cameras, ensuring realtime monitoring and prompt response mechanisms are in place. The successful implementation of AI models will contribute to safer environments and reinforce public trust in surveillance systems.**[17]**

Security cameras are essential for safeguarding various environments, including residential, commercial, and public spaces. However, the increasing interconnectedness of devices and the Internet of Things (IoT) expose security cameras to cyber-attacks, posing risks to privacy, property, and public safety. This project aims to address these emerging threats by leveraging artificial intelligence (AI) models to enhance the security and resilience of surveillance systems. By analyzing network traffic patterns, the project seeks to accurately detect and mitigate attacks on security cameras, ensuring realtime monitoring and prompt response mechanisms are in place**.[18]**

The successful implementation of AI models will contribute to safer environments and reinforce public trust in surveillance systems. For this purpose, the normal dataset will be collected on our end by hardware setup. And the attack dataset will be collected from the official website by Canadian Institute of Cyber Security. There is different AI algorithm that can be used such as Perceptron, Random Forest, AdaBoost and deep neural networks. After this implementation we shall be able to identify if the traffic towards the security camera is normal or it is from an illegal user who is trying to access the control of the camera. Our goal will be to get the maximum results with less false negative and false positive to get the maximum efficiency**.[19]**

## 2.2 What is IoT?

### i. Introduction

The Internet of Things (IoT) is explored in depth in this chapter, with definitions, examples, and discussion of its possible effects across many sectors provided. To gather and share data via the internet, physical devices, sensors, actuators, and software are all part of what is known as the "Internet of Things." Everyday items and big systems alike may benefit from being networked together to form a smart and efficient ecosystem.

### ii. Definition and Components

The Internet of Things (IoT) is a collection of devices and nodes that are able to work together to ease the operations, create more productivity, as well as to give out the right information for better decision making. One can see that there are sensors, actuators, embedded systems, communication protocols, and cloud computing among others that is included in this category. The role of IoT in this phenomenon is to link the digital and physical worlds at the exact moment of occurrence, via data collection and processing, as well as data transmission.

### iii. Data Collection and Communication

Data gathering through sensors and devices is the main foundation of the Internet of Things system considering the situation, the data collection sensors may track different parameters such as temperature, humidity, motion and position. The data flow reports to centralized servers or the cloud by means of wired or wireless networks.

### iv. Connectivity and Network Infrastructure

IoT is dependent on the efficient connection because it permits without interruptions data transfer and continuous communication between devices. Wi-Fi, Bluetooth, RFID, Zigbee, cellular networks, and even satellite communication could be used, consequently. These categories: range, bandwidth, power consumption, and every single IoT application requirements must be taken into consideration so as to determine the most appropriate connection to use.

### v. Applications of IoT

The Internet of Things is a broad spectrum of opportunities, which are already widely used by different sectors. Used in a wide range of applications like energy management, ecosystem monitoring, smart homes, agriculture, healthcare, transportation, manufacturing, and so on. IoT provides for the automatic, optimized and intelligent issuance of commands by linking devices and systems. In medicine, as an example, IoT devices can be adapted to monitor vital signs remotely and aid doctors with instant detection and timely proceedings.

### vi. Challenges and Future Directions

IoT has been perceived with many benefits but they have also created cynicism. Privacy and security of data, building a framework for easy data-sharing, handing the skyrocketing amount of new data, and understanding more about the data model are only a few examples of the problems that neurotech must solve. The ever-continuing process of technology advance, standardization efforts and the use of AI and ML makes the realization of an intelligent globally networked future a possibility.

### vii. Artificial Intelligence

The changing of the cognitive roles often undertaken by man or woman is the goal of this exercise. Conversely, the scope of artificial intelligence as a discipline is not limited to machine learning and deep learning; rather, there are other methods that do not even include learning as one of their techniques. The fact, that AI was the right solution for relatable well defined, logical problems, e.g., playing chess, did not mean that it could create precise rules for the solutions of hard, vague problems, e.g. problems like image classification, speech recognition, and language translation. The favorable trait of a computer science is that it has a logic of causality, which means it makes a system do the things that have the highest probabilities of reaching one's goal. Learning from data and improvements is a section of computer science where computers are able to teach themselves without the support of humans. Traditional methods involve self-discovery and knowledge acquisition through the intake of information in free form such as text, photos or video. Deep learning technologies override spontaneous learning by digesting large quantities of unstructured content.

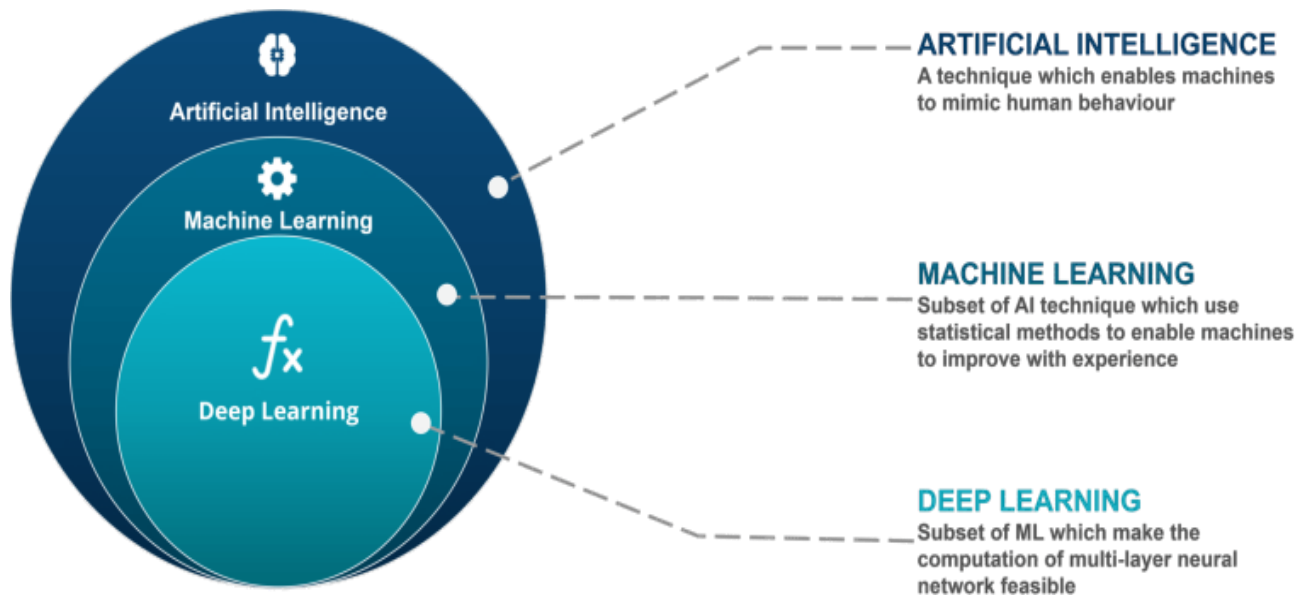

*Figure 2. 1: Artificial Intelligence, Machine Learning, and Deep Learning (Atul, 2018)*

## 2.3 Why AI?

The level of information generated today not only by people but also by technology is practically ten times higher compared to what people can process, grasp, and use to draw a sensible conclusion. The basis of computer science is the prerequisite to any advanced computer programming and the method of progress to higher cognition. AI provided a number practical applications means during the day of running a business can be either more profitable or at a minimum to reduce costs in already established industries.

### i. Machine Learning

Machine learning is AI type that offers machines the opportunity to learn from their mistakes they made by themselves and become a better version of themselves without writing any secret treaties. Machine learning narrows down this goal by giving machines a capability of self-learning and the ability to create their own programs. Such freedom makes machines learn and behave just like people. Mostly it is done by AI as a matter of fact with this phrase no particular coding is required. The instruction is led by technical expertise, and the machines have helped it to be streamlined further. The knowledge has been provided to machines in a high quality and different approaches have been applied to build model classes for the same metric system of units as the ones used to train computers. Whether there has to conduct a simple operation or a complicated task determines which template is selected.

**Machine Learning Techniques**

Two categories of machine learning exist:

- Machine learning under Supervision
- Automatic Machine Learning

## ii.     Reinforcement Learning

We shall not stop there and look into the various machine learning strategies available to improve on the training of a machine learning model. The state of the economy at the time can be shown by the pattern of savings., and the buyers and sellers are impacted by it. A number of elements will determine how much a house will finally amount to. The number of bedrooms, bathrooms, and the location are among some of those requirements. Country areas … including costs are lower than those of cities. Due to the fact that the cost of the property goes up when things like access of highways, shopping centers, supermarkets, places of jobs, top educational institutions, etc., firms in the property business tried to forecast property values in the past manually. The company uses a qualified executive group to set the price grade of any real estate property. To arrive at the right price, they look back at their past sales history. Nevertheless, it is worth mentioning that there is a 25% inaccuracy in the given forecast. Consequences of such move are detrimental to the businesses and budgets. Consequently, A number of algorithms have emerged to help investors with predicting the prices of the real estate properties. The introduction idea of the housing forecasting system seems from Rick Siow and Si, Fei, Lu. The setting of an approximated model that could accurately value properties according to other factors was the system's main purpose.

Deep learning is generally regarded as the most straightforward method for modifying predictive analytics. Deep learning algorithms are piled in a hierarchy of rising complexity and abstraction, as opposed to older machine learning algorithms, which are linear.

The steps required for machine learning model training can be summed up as follows:

   i. Data preparation
   ii. Division of the dataset
   iii. Deciding on a model
   iv. Outlining the optimization algorithm and loss function
   v. Model Training
   vi. Assessing the model.   vii.   Modifying the model.   viii.   Model Testing

## 3.1 Methodology

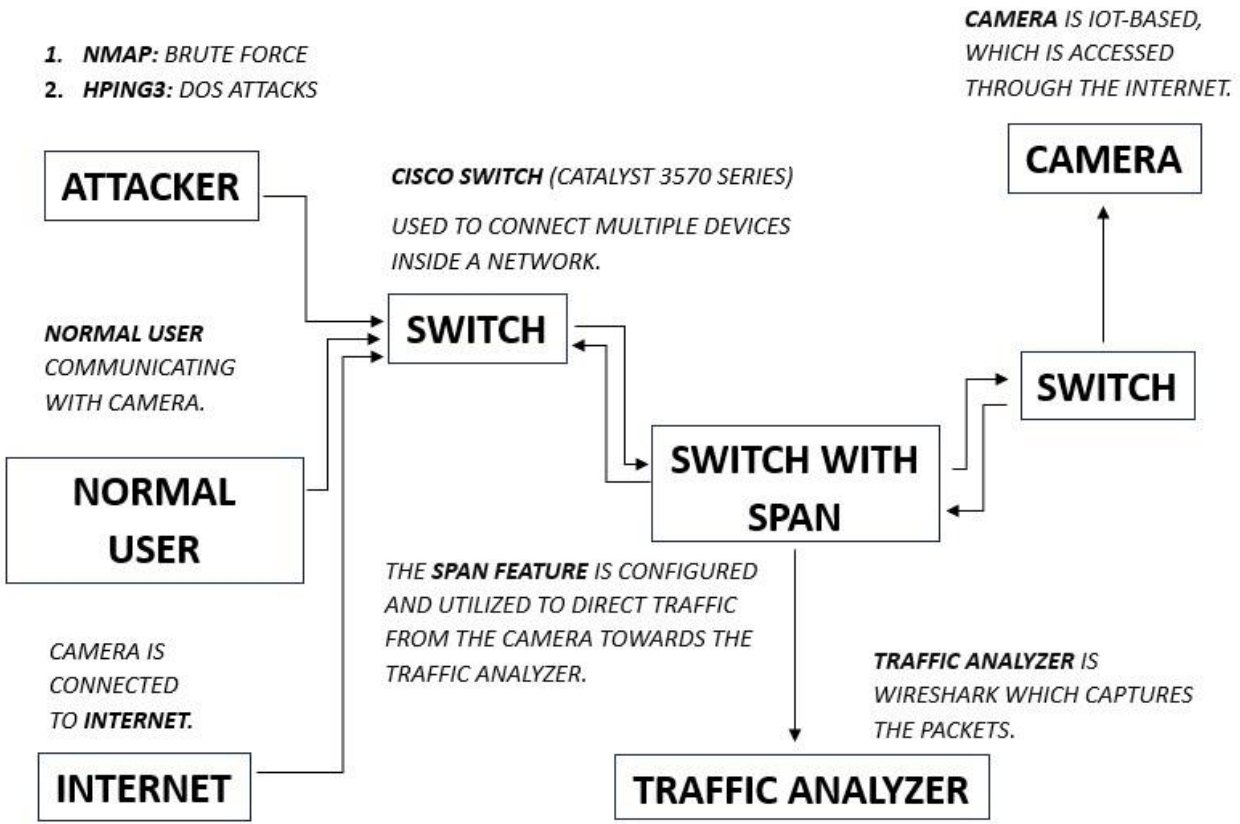

*Figure 3. 1: Block Diagram*

With the growing concern of a distributed denial of service attack exercised by an unidentified hacker employing Nmap brute force and Hping tools, a network traffic analysis strategy to be used as the counter measure is available. Recognizing that the situation of immediate identification and disabling such threats is critical, the supporting network infrastructure for the Cisco Catalyst switch acting as a gateway to users and an access point for various devices is being implemented with advanced controlling and analysis features.

The Switched Port Analyzer (SPAN) port has been configured to strengthen the security on the Cisco switch. This SPAN port acts as a watchful guardian of the invader's network, by capturing and duplicating of all packets meant for the compromised camera. It is by the use of a packet sniffer to intercept these packets, and then with the aid of a designated network traffic analyzer like Wireshark, that they are minutely scanned. Wireshark is arguably the most popular and powerful of all the network packet analysis tools. Unlike most of its counterparts, it is designed to capture and dissect packets, and thus it has the ability to meticulously log every piece of data crossing the network, providing valuable clues about the nature and origin of the attack.

The investigation, however, does not end here. To make the defensive measures more robust in their execution, command-line Tshark, which is a counterpart to Wireshark, is also deployed. This powerful tool is able to dig deep into the rectal info coming from the captured packets and eventually reveal hidden temporal patterns and abnormalities that may suggest malicious activity. These features taken out provide ML model as a base for it assisting in detecting and labeling any DoS attacks.

## 3.2 Types of Attack

In this project we have focused mainly on the ***Denial of Service*** and ***Brute Force Attacks***. Being the most dangerous types of attack for cameras, because lack of awareness on how to update these devices.

**Denial of Service:**

These are the types of attacks that are considered to be dangerous. They work by sending continuous packets or requests to the camera until it breaks and disconnects itself from the all the devices and becomes inaccessible. Since these attacks are easy to pull off because of the material on the cyber security. And most of the users are not aware how to make these devices more secure.

- TCP SYN Flood Attack: This attack exploits the TCP three-way handshake process. In a SYN flood, the attacker sends a large number of SYN (synchronize) packets to the camera's network interface, each appearing to be a legitimate connection request. The camera responds with SYN-ACK packets, but the attacker never completes the handshake by sending the final ACK packet.
- UDP Flood Attack: In a UDP flood, the attacker sends a large number of UDP packets to the camera's network interface, typically to random or specified ports. Because UDP is connectionless, the camera must process each incoming packet. This flood of packets can overwhelm the camera's processing capability, causing it to slow down or become completely unresponsive to legitimate traffic.

**Brute Force:**

These types of attacks work by continuously trying for different username and passwords until the dictionaries is completed. And there are thousands of dictionaries out there. And they are easily accessible to attackers. So, these attacks can be conducted with simple set of steps.

We are tackling these attacks by making such a prototype that can capture all the traffic in the network directed towards the camera. And then analyzing the model and implementing machine learning model to detect and classify the attack directed towards the camera.

## 3.3 Software and Hardware Components Overview

### i. Google Collaboratory

Google Collaboratory, or Google Collab, is a cloud-based virtual environment that can be used to write and run Python code provided by the Google company. It gives Jupiter Notebook- like interface. Free to use computing resources, including CPU and GPU are included, so it can be used when running machine learning algorithms and models.

### ii. Python

It is a widely-used programming language in the field of data science and machine learning, was used in this project. Python offers a rich ecosystem of libraries and frameworks that facilitate data manipulation, analysis, and modeling. It offers an interface that is simple to navigate and an extensive assortment of utilities that hasten the progress of development.

For this undertaking, we employed Google Collab as the coding environment owing to its ease of use and the abundance of resources it offers. It seamlessly integrates with Google Drive, facilitating effortless access and sharing of datasets and code files. Furthermore, Collab boasts GPU support, which can notably enhance the speed of neural network training and enhance overall efficiency.

### iii.     Kali-Linux

Kali Linux, the renowned ethical hacking distribution, isn't just a software; it's a digital arsenal designed for cybersecurity professionals. Built on Debian's stable foundation, Kali packs a punch with over 600 pre-installed tools specifically crafted for penetration testing, security research, forensics analysis, and vulnerability assessments. This vast toolkit empowers security experts to simulate real-world attack scenarios, identify weaknesses in systems, and propose effective security measures.

### iv.     Nmap

A well-known open-source software, Nmap **[20]**, more commonly known as "Network Mapper," is a utility that allows administrators to both identify and assess target networks. The main purpose of this scanner is to sweep through networks to uncover web servers, hosts, and programs running on these servers by sending packets and passing their answers through the analysis. In addition to that, Nmap features an engine for scripts called NSE (Nmap Scripting Engine) which was designed for allowing users to write and share their scripts in accordance with a set of network related tasks like vulnerability detection and service enumeration. Nmap in its original form does not provide any brute force attack options however, it makes perfect sense to utilize this software in conjunction with external tools or scripts of this kind since it forms a major part of most security audits and ethical penetration attempts.

### v.     Hydra

Hydra **[25]** is known as a password breaking code that is implemented in all professional pen testing and security auditing workshops. It supports a range of protocols which enable the brute force attacks like SSH, FTP, Telnet HTTP(S), SMB, and others. The dictionary attack feature is one of the main characteristics of this software. It will run through all the possible identity and password combinations from the words in the online dictionary or a given wordlist. Hydra's ability to be extremely customizable as well as fully multithreaded, a simple truth, makes it one of the most effective tools for attempting to guess passwords across multiple service providers simultaneously. Providing integration for several protocols as well as the capability to work with customized modules, Hydra offers great flexibility for testing of application security vulnerabilities and also injection of flaws in authentication mechanisms.

## vi. Hping

Hping **[21]** is a command line tool that is mostly utilized to gather information from networks and test networks. It generates ICMP, UDP, TCP, and RAW-IP packets manually and observes any reply from the points targeted. Hping can be used for the diverse tasks, for example, network infrastructure performance testing and firewall analysis, but it can also be a tool for brute force for securing IoT Devices that are Internet of Things has especially the cameras as a target. Its flexibility and scripting features allow users to craft reconfigurable packets by which they can stage many attack types, for instance SYN flooding and serial port scanning. On the one hand, it is important to acknowledge that most likely Hping will be used by somebody who crafts a packet in order to exploit the camera or its misconfigurations rather than by somebody who will try to guess the passwords directly. Although Hping is not widely known in the context of password cracking as some other tools like Hydra are, it is definitely a popular tool for testing and security assessments, offering capabilities for more than just the simple port scanning.

## vii. Hardware Equipment's used

 i. IoT Cameras
 ii. Switch with Span feature
 iii. Router with port forwarding

IoT Cameras:

The EZVIZ C6N is a smart Wi-Fi pan and tilt camera offering comprehensive home security with features like 360° coverage, smart tracking, and clear night vision. Its 1080p resolution ensures detailed visuals.

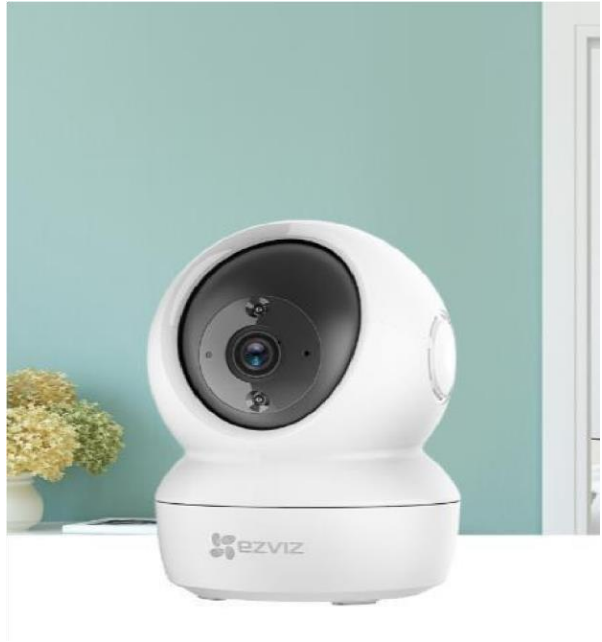

*Figure 3. 2: IoT Camera* Switch

Catalyst 3570S:

The Cisco Catalyst 3570S switch is equipped with a powerful feature known as Switch Port Analyzer (SPAN), which plays a vital role in network analysis. As it can act as a network tap for capturing the traffic directed to a specific port. The Span provides the insights to the administrator about the flow of traffic. It also works by specifying the particular port destination. Furthermore, Span enables administrators to conduct in-depth analysis of network traffic patterns, protocols, and application behavior. By capturing packets in real-time or for later analysis, administrators can gain valuable insights into the operation of their network infrastructure and make informed decisions to optimize performance and enhance security

Router 1871:

The Cisco 1871 router serves as a crucial component in establishing a network infrastructure that enables seamless connectivity and accessibility for cameras to the outside network. Its advanced features and capabilities make it an ideal choice for facilitating secure communication and data transmission between cameras and external networks.

One of the primary functions of the Cisco 1871 router in this scenario is to provide network connectivity for the cameras, allowing them to transmit video footage and data to external devices or servers. The router acts as a gateway, managing the flow of traffic between the local camera network and the broader external network, which could include the internet or other connected networks.

## 3.4 Project Workflow

### 3.4.1 Extracted Features

- <u>Pseudo Code</u>:

Importing the libraries (pandas, NumPy, scikit-learn, pickle)  Function feature extraction:

    Using Tshark command to extract features (ttl, ip_len……)
    Starting subprocess and adding the Tshark command
    Communicating the process through out any error in the process
    Def Capture:  To start live capturing
        Def interface
        Specify capture duration
        Writing Tshark, cmd and specifying interface duration
    Starting the subprocess to execute the Tshark commands.  ***Main:***

Starts Capturing for each label and storing them in file.

Extracting features in the form of Stored file, them converting it into CSV file.

Labeling the data in each file.

Applying trained models to get Accuracy, Precision, Recall and F1 Score. Printing the results for each of them.

- Features:

Following are the features extracted from the given Pcap files using Tshark:

| Ip.scr | Ip.ttl | Tcp.time_relative | Udp.stream |
|---|---|---|---|
| Ip.dst | Ip.id | Tcp.analysis.rtt | Udp.length |
| Ip.proto | Ip.hdr_len | Tcp.window.size | http.request.method |
| Eth.src | Ip.len | Tcp.hdr_len | http.response.code |
| Eth.dst | Ip.flags.df | Tcp.len | http.content_length |
| Ipv6.src | Tcp.stream | Udp.scrport | Frame.time |
| Ipv6.dst | Tcp.time_delta | Udp.dstport | Frame.len |

As the features are transported to machine learning algorithm. Consequently, this type of information is only given in numbers. Numeric values only will be considered (ignore non numeric). As they are made from different types of information, the result of the research will be different. That is why the implementation of the model is seriously hindered.

## 3.4.2 Flow-Chart

*TRAINING MODEL*

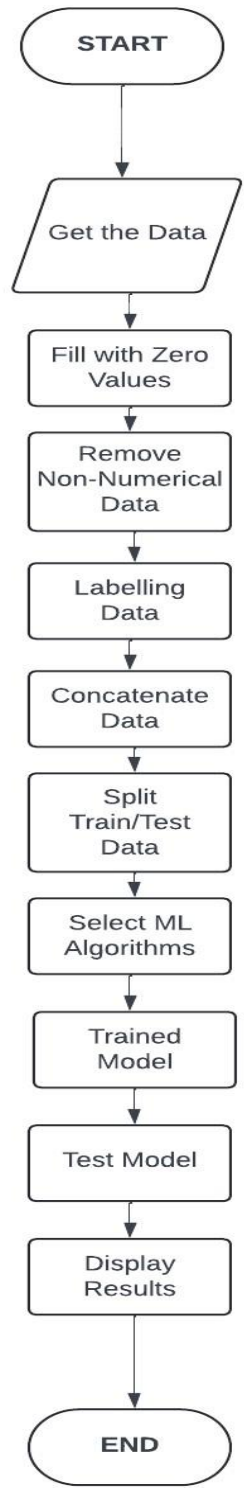

*Figure 3. 3: Training Model*

This chart is a flow chart which shows the sequence of steps involved in training the machine learning models:

Start: We start with the initiation of the aforementioned data preparation pipeline primarily as we indulge into the process of training a machine learning model for the intended task.

Get Data: At this critical point, the data which is indispensable for the development of a machine learning model is accurately sourced from the relevant parties. This information is the basic component that offers the model the capability to make forecasts.

Split Data: Then the data entered, it has an important parting process, the data is divided into two-part training data and test data. Com positions are usually around 70% for training purposes, and the remaining 30% are set aside for checking out how the performance of the model turns.

Fill with Zero Values: The process of tackling any vacancies that might be there in the training data precedes accordingly, where the process of filling the missing values with zeros becomes meticulous. Thus, this step ensures that the data set is not having any blank spaces that could get in the way of the model's trainings.

Remove Non-numeric Data: Primarily, the requirement of many machine learning algorithms is working on the numerical data and, therefore, non-numeric data has to be systematically eliminated from the train data set. This helps with selecting and using only those data types that are compatible with the model training process in the next stages.

Concatenate Data: Then, the cleanup stage of doing the missing value filling and nonnumeric data elimination is followed by joining the ingredients of the datasets together before concatenation into one unified whole. This amalgamation is the precursor to the training of the machine learning software, which is the next step to follow.

Train Data: This highly noted dataset of the machine learning model, to be particular, the Random Forest model, then undertakes the very extensive training process. This part of the model is where it learns deep and complex patterns involved in the data which facilitates it to predict not known data.

Test Model: With the training phase successfully done, the level of effectiveness of the well-trained model is thoroughly assessed by the use of the unseen testing dataset. After this step the model will be tested on the data that it has not been trained on, and thus it signifies if the model is robust and if it generalizes well.

Results: At the final stage of the workflow are the findings from the testing phase, it is the results after being examined. Those results are usually composed of a robust variety of performance metrics, namely accuracy, precision, recall and F1-score. Those metrics are of great interest to the model's stakeholders for evaluating the model's success in the intended task.

**TESTING MODEL**

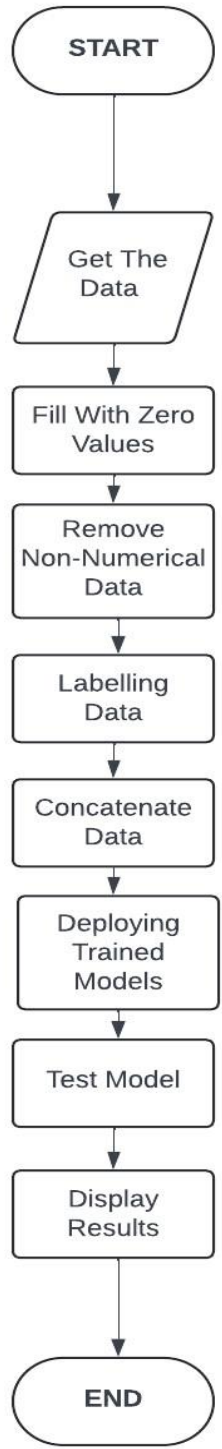

*Figure 3. 4: Deploying ML Model*

Start: This is the initiating point of the data preparation pipeline for training a machine learning model.

Get Data: In this step, data that's relevant to the machine learning task is collected from various sources. This data serves as the foundation for the model's ability to make predictions.

Split Data: The data is then divided into two sets: training data and test data. A common split is 70% for training and 30% for testing. The training data is used to train the model, while the test data is used to evaluate the model's performance.

Fill with Zero Values: This step addresses missing values in the training data. Here, missing entries are filled with zeros. This is a common approach to handle missing data, but there are other techniques that can be employed depending on the data and the machine learning model.

Remove Non-Numerical Data: Non-numerical data is then eliminated from the training data. Machine learning models typically work best with numerical data. There are techniques to convert categorical data into numerical data.

Labeling Data: In some cases, the data may need to be labeled. For instance, if the task is image classification, the images would need to be labelled with the corresponding object category.

Concatenate Data: After the data is preprocessed, it may need to be combined from multiple sources. This process is known as concatenation.

Deploying Trained Models: Once the data is prepared, it can be used to train a machine learning model. The trained model can then be deployed for real-world applications.

Test Model: The model's performance is assessed using the test data that was set aside earlier. This assessment helps determine how well the model generalizes to unseen data.

Results: Finally, the results of the model's performance on the test data are obtained.

## 4.1 Results (Training)

### 4.1.1 Getting Data

The data was first taken from ***Canadian Institute of Cyber Security (CIC).* [26]** And trained on the data set with different variation but when tested on live scenario gave a very poor accuracy which was less than 60%. But when the data was re-trained on generated dataset on the camera with varying dataset a few of them performed really well. The data was generated using different variations with different sizes and utilizing different options available to make the attack more effective. So, the system is hardware dependent and this marks the limitation of the project.

## 4.1.2 Simulation Settings

- **Simulation Settings for Attack Generation**

The attacks were generated using Kali Linux with a tool called hping3 and Nmap. Following the command used

*For Tcp Dos attack*:

   hping3 -S -c <Number of packets to send> <target Ip> -p 80 –faster *For Udp Dos attack*:

   Hping3 -2 -c <Number of packets to send> <target IP> -p 50160 –faster *For Brute force attack*:

   Nmap –script rtsp-url-brute -p 554 <ip>

For multiple brute force attempts the rtsp-script in Nmap was ran multiple times to properly generate the data and train a model on this attack.

- **Simulation Settings for ML model:**

Following were the parameters used to train ML Models:

*Random Forest Classifier*:
RandomForestClassifier(n_estimators=60, max_features="sqrt", random_state=101)
*Ada-Boost*:
   adaboost_classifier = AdaBoostClassifier()

*Log Regression:*   logreg_model = LogisticRegression(random_state=42) *Navie Bayes*:
   GaussianNB()

*Perceptron*:
   # Standardize features (important for perceptron) scaler = StandardScaler()
   X_train_scaled = scaler.fit_transform(X_train)
   X_test_scaled = scaler.transform(X_test)

```
# Create a perceptron model
perceptron_model = Perceptron(random_state=42)
```

### 4.1.3 Training

The machine learning models were trained on 3 different Data Sets. The data was trained to see how the model will respond in training as well as in testing. Since there are **4 Labels in the Dataset** mainly: ***TCP flood, UDP flood, Brute force and Normal traffic.***

*90k per Label:*

For the first set of models, the data size was kept 90k Packets per label and the total size on which the models were trained is 360k Data Points.

| Trained Models | Accuracy |
|---|---|
| Random Forest | >90% |
| Ada-Boost | >50% |
| Log Regression | >80% |
| Naïve Bayes | >80% |
| Perceptron | >90% |

*500k per Label:*

For the second set of models, the data size was kept 500k Packets per label, summing up-to 2M Data Points. Each model was trained on these points, the table below shows the accuracy of the trained models.

| Trained Models | Accuracy |
|---|---|

| Trained Models | Accuracy |
|---|---|
| Random Forest | >90% |
| Ada-Boost | >90% |
| Log Regression | >90% |
| Naïve Bayes | >70% |
| Perceptron | >90% |

### Imbalanced Trained Model:

For the third set, the data size was irregular to see the accuracy of the models during training and testing as well.

| Trained Models | Accuracy |
|---|---|
| Random Forest | >90% |
| Ada-Boost | >60% |
| Log Regression | >90% |
| Naïve Bayes | >70% |
| Perceptron | >97% |

Few of the models achieved the accuracy of 99% or even 100% percent. So, to see if the model was overfitting. The data was augmented by adding, subtracting different values from the data and trying to visualize data from different angles and then training and testing to see if it really overfits or not. But the results after augmentation were also the same.

## 4.2 Results (Testing)

The models were tested on the data size of **330K Packets** *(Almost).*

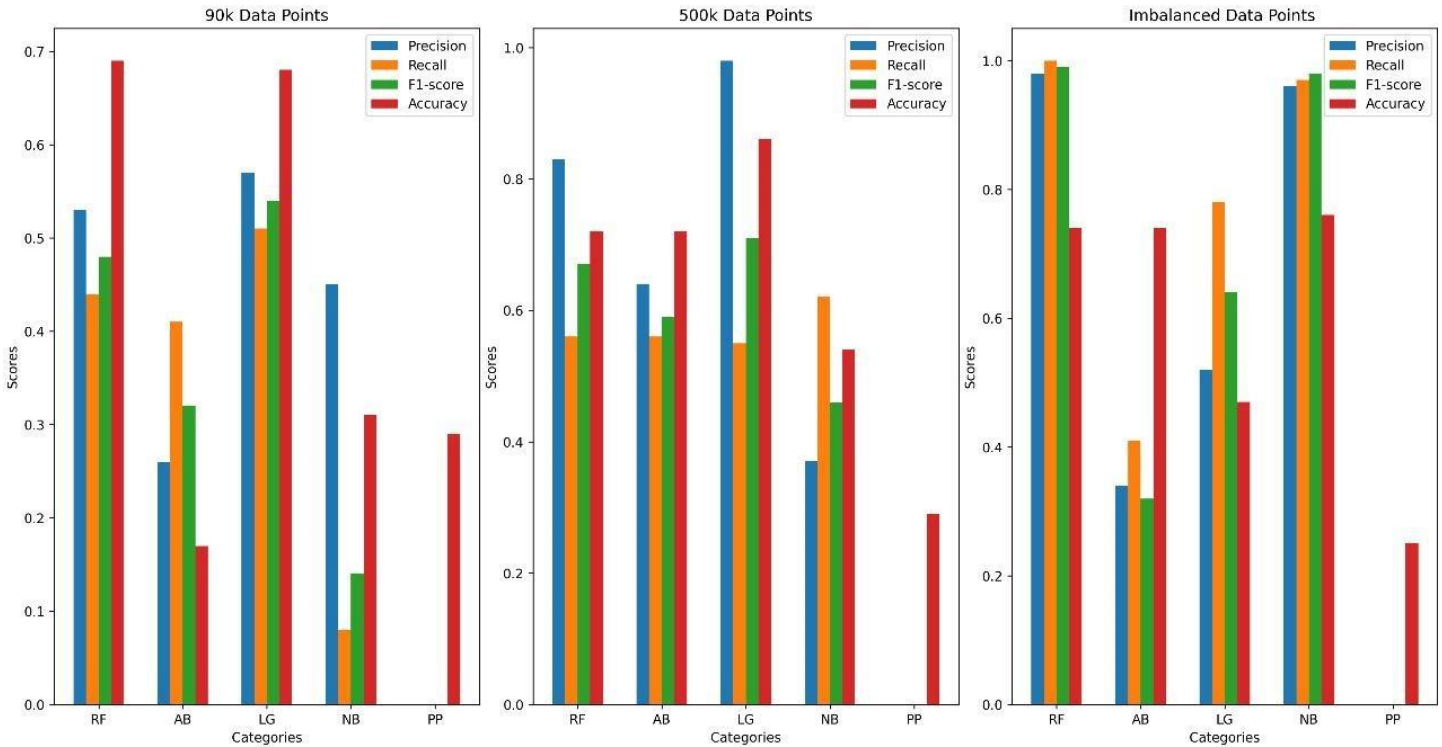

*Figure 4. 1: Result on Data size of 330K packets (Almost)*

Among the trained models, the one trained on **90k Data Points** generally exhibits the highest precision, recall, and F1 score, particularly for the Random Forest model, which consistently surpasses 0.9 for these metrics. However, the Perceptron model shows notably lower accuracy, around 0.3.

In the **Imbalanced Trained Model**, Naive Bayes stands out with the highest accuracy of 0.9, while Perceptron remains at the lowest accuracy of 0.3.

For the model trained on **500k Data Points**, both Random Forest and AdaBoost demonstrate precision, recall, and F1 scores ranging from 0.3 to 0.4. Again, Perceptron performs poorly with an accuracy of 0.3. Notably, Naive Bayes exhibits the lowest F1score of 0.3 in this scenario.

The models were tested on the data size of **330K Packets.**

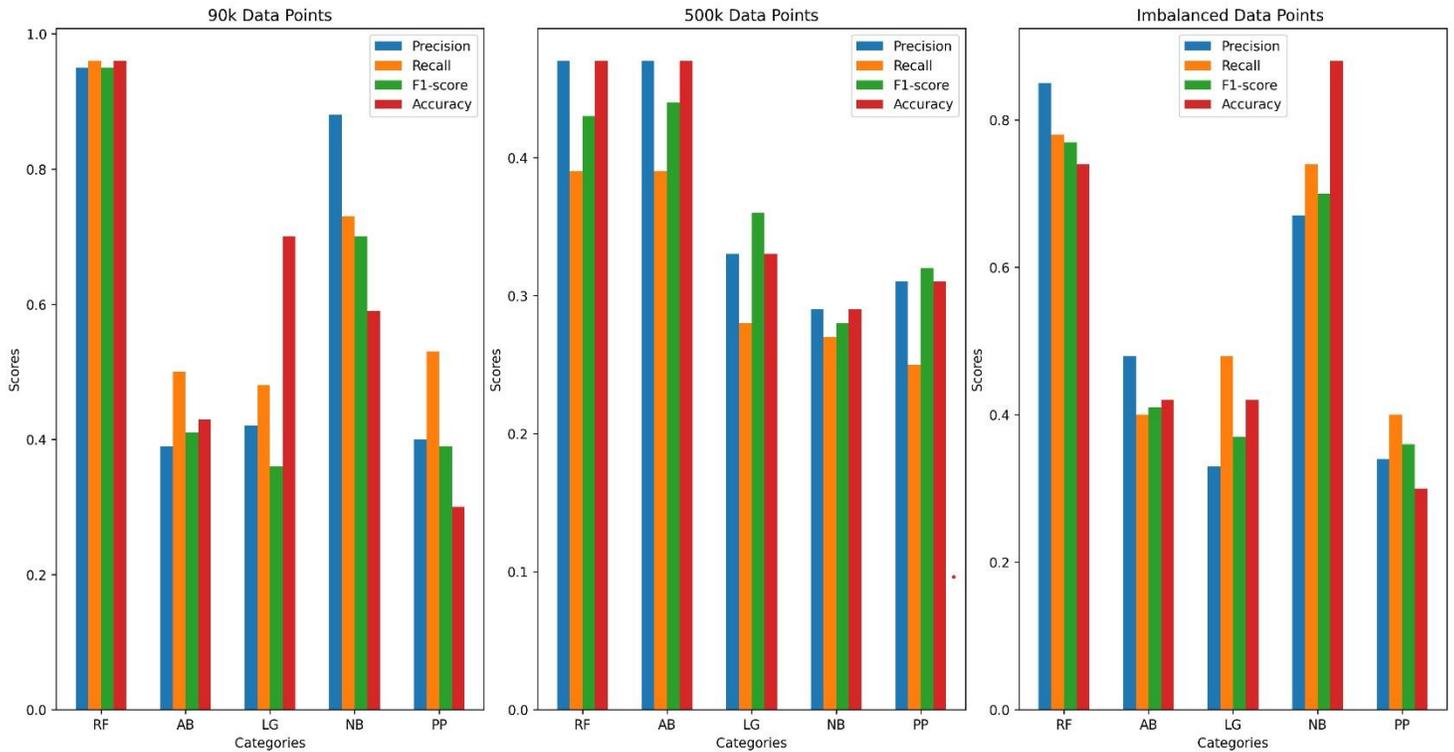

*Figure 4. 2: Result on Data size of 330K packets*

Among trained models, the Random Forest model trained on **90k Data Points** consistently achieves the highest precision, recall, and F1 score, all surpassing 0.9. Conversely, the Perceptron model exhibits the lowest accuracy at 0.3 across all scenarios.

In the **Imbalanced Dataset**, Naive Bayes records the highest accuracy at 0.9, while Perceptron maintains the lowest accuracy of 0.3.

In the case of the **500k Data Points Model**, both Random Forest and AdaBoost models yield precision, recall, and F1 scores ranging from 0.3 to 0.4. Once again, Perceptron displays the lowest accuracy of 0.3. Notably, Naive Bayes also exhibits the lowest F1score of 0.3 in this scenario.

The models were tested on the data size of **400K Packets**.

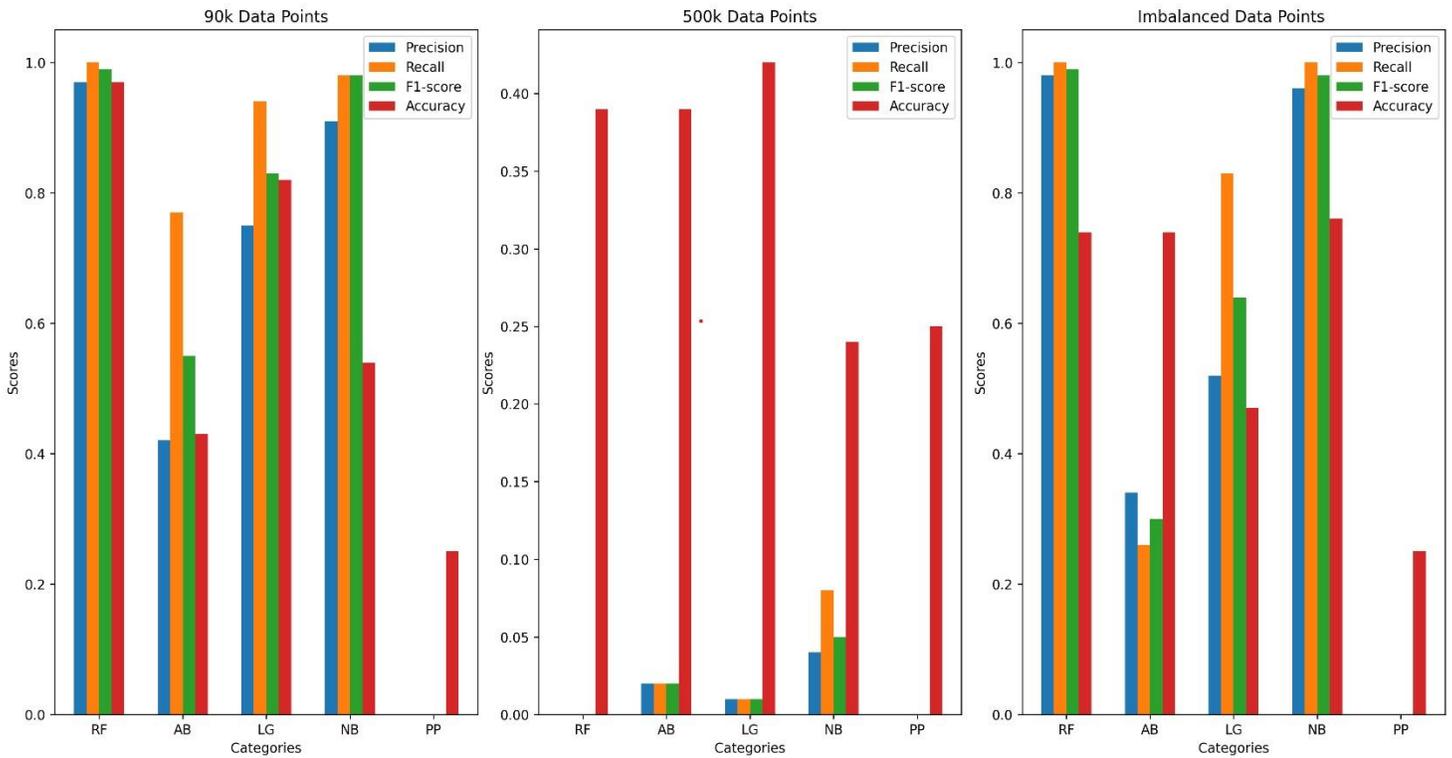

*Figure 4. 3: Result on Data size of 400K packets*

From the provided graphs, it's evident that the Random Forest model trained on **90k Data Points** consistently maintains high precision, recall, and F1 score, while Perceptron's performance notably declines, reaching zero for recall and precision, with an accuracy of 0.3.

In the **500k Data Points Model**, Logistic Regression achieves an accuracy of 0.5, while both Random Forest and Perceptron show zero precision, recall, and F1 score.

In the **Imbalanced Dataset**, both Logistic Regression and Random Forest achieve an accuracy of 0.6, with Random Forest exhibiting the highest precision, recall, and F1 score, alongside Naive Bayes.

The models were tested on the data size of **1.3M Packets.**

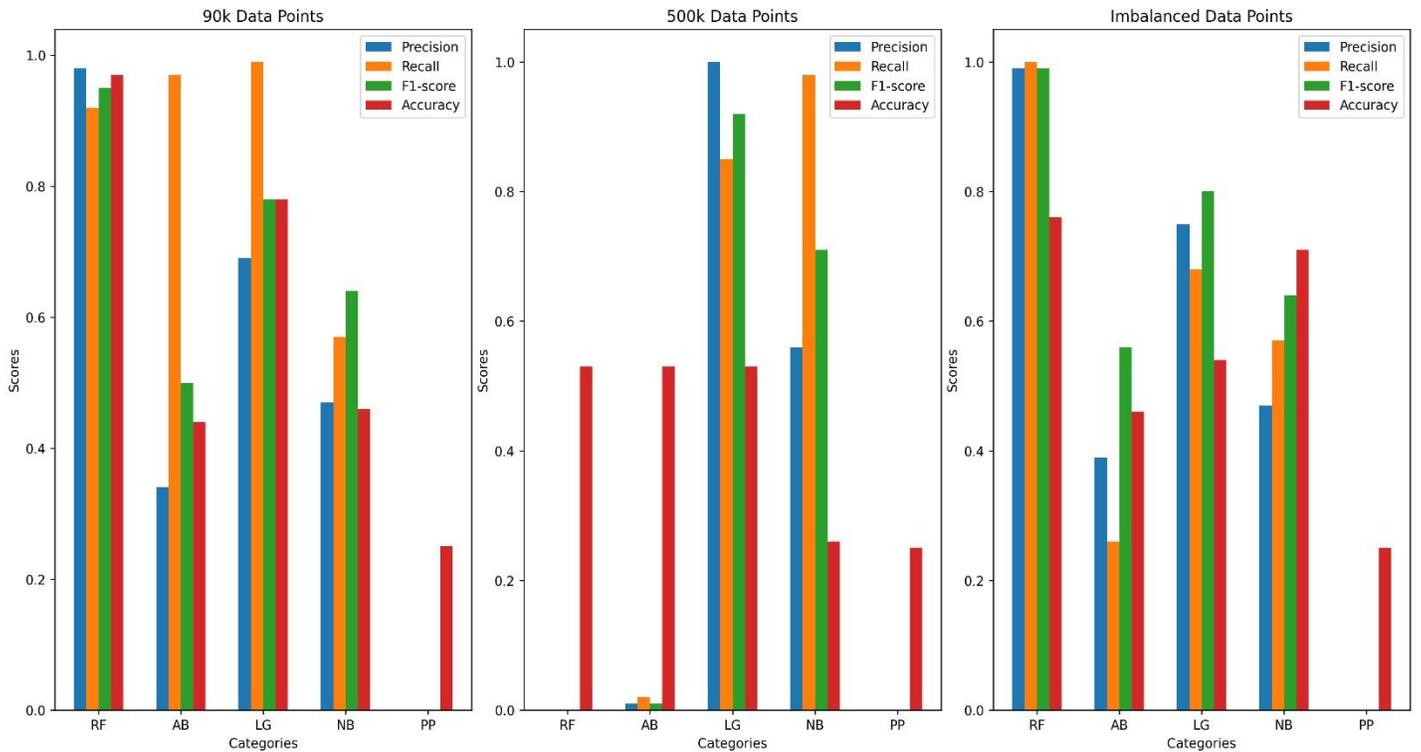

*Figure 4. 4: Result on Data size of 1.5M packets*

From the provided graphs, it's evident that the Random Forest model trained on **90k Data Points** consistently maintains its performance, while Perceptron's precision and recall drop to zero, resulting in the lowest accuracy of 0.3.

In the **500k Data Points Model**, Logistic Regression achieves an accuracy of 0.5, while both Random Forest and Perceptron show zero precision, recall, and F1 score. In the **Imbalanced Dataset**, Logistic Regression achieves an accuracy of 0.6, and Random Forest maintains the highest precision, recall, and F1 score among all models.

## 5.1 Conclusion

The project "Detection and Classification of Attacks on IoT Cameras" was a comprehensive endeavor aimed at enhancing the security of IoT camera systems. It involved an intricate process of feature extraction from Wireshark files, meticulously implemented using various research papers and a selection of Python tools including DPKT, PyShark, Nfstream, and Tshark. Among these tools, Tshark emerged as the standout performer, recognized for its efficiency and effectiveness, particularly when operating within Windows environments. Leveraging the capabilities of Tshark, the project team successfully extracted pertinent features essential for identifying and classifying potential attacks on IoT cameras. This meticulous feature extraction process laid the groundwork for subsequent stages of the project, setting the stage for the evaluation and implementation of machine learning models.

Following the feature extraction phase, the project progressed to training and testing various machine learning models using the extracted features. Through rigorous experimentation, the team evaluated the performance of different models across metrics such as accuracy, precision, recall, and F1-score. Notably, the Random Forest model emerged as the top performer in terms of accuracy, consistently achieving remarkable results when trained on a dataset consisting of 90k packets per label. With accuracy levels consistently hovering between 0.95 to 0.96, the Random Forest model showcased its efficacy in accurately detecting and classifying potential attacks on IoT cameras. This achievement underscores the significance of employing sophisticated machine learning techniques in bolstering the security infrastructure of IoT devices, particularly in domains as critical as surveillance and monitoring systems.

## 6.1 Exploring Enhanced Data Collection and Integration

Future research will involve a development of better data collecting and integrating approaches. In addition to analytical tools, new data sources have been made accessible from technological development, such as IOT devices, social media, and geolocation data also. The potential of capturing the real estate market just got better with the incorporation of these vast data sources, which in turn may be able to generate even more precise forecasts and insights as well. The major aspect in this is the opening of the access to reliable gathering procedures, preservation of the data and forming to well-linked data sources.

## 6.2 Advancing Analytics and Machine Learning Techniques

The challenge for future research is implementation of complex analytics and machine learning methods which will help improve accuracy of predictions. Deep learning models, the Natural Language Processing (NLP), and computer vision algorithms might all be applied to extract high-quality information from different chaotic data sources such as real estate photos, description of the housing listings, and user reviews. Employing these strategies can assist in the prediction of market trends accurately, find investment advantages and improve property matrices. Implementation of explainable AI techniques can be an instrument which can lead to transparency and interpretation.

## 6.3 Collaboration and Partnerships with Industry

The extent to which the project succeeds and influences other businesses in the real estate sector comes with collaboration with industry players consisting of real estate firms, property developers and the government. The importance should be considered in coordinating and integrating resources to obtain credible data sources, inspect models, and get consultation and facts from real life. Through the interaction with the experts, the innovation can be designed and invented in a way that allows it to meet the needs of the different sectors and industries. The adoption is likely to be high, and the return on investments is foreseen.